\documentclass[prl,reprint,superscriptaddress,showpacs,preprintnumbers,nobibnotes,amsmath,amssymb, aps,longbibliography]{revtex4-2}

\usepackage[T1]{fontenc}
\usepackage[colorlinks=true,citecolor=blue,linkcolor=blue]{hyperref}
\usepackage{cleveref}
\usepackage{booktabs}
\usepackage{graphicx}
\usepackage{siunitx}
\usepackage{enumitem}
\usepackage{multirow}
\usepackage{pifont}
\usepackage[dvipsnames]{xcolor}
\usepackage[normalem]{ulem}
\usepackage{xspace}
\usepackage{lineno}

\begin{document}

\title{KM3NeT Constraint on Lorentz-Violating Superluminal Neutrino Velocity}
\begin{abstract}
Lorentz invariance is a fundamental symmetry of spacetime and foundational to modern physics.
One of its most important consequences is the constancy of the speed of light.
This invariance, together with the geometry of spacetime, implies that no particle can move faster than the speed of light.
In this article, we present the most stringent neutrino-based test of this prediction, using the highest energy neutrino ever detected to date, KM3-230213A.
The arrival of this event, with an energy of $220^{+570}_{-110}\,\text{PeV}$, sets a constraint on $\delta \equiv c_\nu^2-1 < 4\times10^{-22}$.
\end{abstract}


\collaboration{The KM3NeT Collaboration}


\author{O.~Adriani}
\affiliation{INFN, Sezione di Firenze, via Sansone 1, Sesto Fiorentino, 50019 Italy}
\affiliation{Universit{\`a} di Firenze, Dipartimento di Fisica e Astronomia, via Sansone 1, Sesto Fiorentino, 50019 Italy}
\author{S.~Aiello}
\affiliation{INFN, Sezione di Catania, (INFN-CT) Via Santa Sofia 64, Catania, 95123 Italy}
\author{A.~Albert}
\affiliation{Universit{\'e}~de~Strasbourg,~CNRS,~IPHC~UMR~7178,~F-67000~Strasbourg,~France}
\affiliation{Universit{\'e} de Haute Alsace, rue des Fr{\`e}res Lumi{\`e}re, 68093 Mulhouse Cedex, France}
\author{A.\,R.~Alhebsi}
\affiliation{Khalifa University of Science and Technology, Department of Physics, PO Box 127788, Abu Dhabi,   United Arab Emirates}
\author{M.~Alshamsi}
\affiliation{Aix~Marseille~Univ,~CNRS/IN2P3,~CPPM,~Marseille,~France}
\author{S.~Alves~Garre}
\affiliation{IFIC - Instituto de F{\'\i}sica Corpuscular (CSIC - Universitat de Val{\`e}ncia), c/Catedr{\'a}tico Jos{\'e} Beltr{\'a}n, 2, 46980 Paterna, Valencia, Spain}
\author{A.~Ambrosone}
\affiliation{Universit{\`a} di Napoli ``Federico II'', Dip. Scienze Fisiche ``E. Pancini'', Complesso Universitario di Monte S. Angelo, Via Cintia ed. G, Napoli, 80126 Italy}
\affiliation{INFN, Sezione di Napoli, Complesso Universitario di Monte S. Angelo, Via Cintia ed. G, Napoli, 80126 Italy}
\author{F.~Ameli}
\affiliation{INFN, Sezione di Roma, Piazzale Aldo Moro 2, Roma, 00185 Italy}
\author{M.~Andre}
\affiliation{Universitat Polit{\`e}cnica de Catalunya, Laboratori d'Aplicacions Bioac{\'u}stiques, Centre Tecnol{\`o}gic de Vilanova i la Geltr{\'u}, Avda. Rambla Exposici{\'o}, s/n, Vilanova i la Geltr{\'u}, 08800 Spain}
\author{L.~Aphecetche}
\affiliation{Subatech, IMT Atlantique, IN2P3-CNRS, Nantes Universit{\'e}, 4 rue Alfred Kastler - La Chantrerie, Nantes, BP 20722 44307 France}
% ...


\maketitle

\section{Introduction}

Lorentz invariance, which states that physical phenomena look the same for all inertial observers, is a key component underlying the Standard Model of particle physics.
Lorentz invariance \textit{violation} (LIV), while so far unobserved, is predicted by models of quantum gravity~\cite{Li:2023wlo} which are parametrized by effective field theories such as the Standard Model Extension~\cite{Colladay:1996iz,Kostelecky:2003fs,Colladay:1998fq,Kostelecky:2008ts}.

Since an observation of LIV would provide compelling evidence of such new physics, it has experimentally been tested in various ways: for example, using electronic transitions~\cite{Hohensee:2013cya}, gamma-ray bursts~\cite{Chen:2024zuj}, high-energy neutrino oscillations~\cite{IceCube:2017qyp}, and top quark production at colliders~\cite{CMS:2024rcv}.

Lorentz invariance also predicts the constancy of the speed of light and therefore, that the speed of light in vacuum is the upper bound on the speed of any massive particle; if one is found to be superluminal, that would unambiguously indicate LIV.
As such, superluminality has been probed with particles such as electrons and cosmic rays~\cite{Jacobson:2002ye,Coleman:1997xq,Stecker:2013jfa,Coleman:1998ti}.
Neutrinos, as the lightest known massive particles, can provide another probe of LIV as they propagate.
Several experimental searches for superluminal neutrino propagation have been performed, for instance, at OPERA~\cite{OPERA:2012der} and MINOS~\cite{MINOS:2007cyk,MINOS:2015iks}; while conclusive evidences of superluminal propagation, and therefore LIV, have not been observed, limits have been set.

Superluminal propagation is characterized~\cite{Cohen:2011hx,Huo:2011ve} by a parameter
\begin{equation*}
    \delta \equiv c_\nu^2 - 1,
\end{equation*}
where $c_\nu$ is the neutrino speed in units of the speed of light. 
A superluminal neutrino rapidly loses energy primarily via the process of pair emission of electrons $\nu \rightarrow \nu + e^+ + e^-$~\cite{Cohen:2011hx,Huo:2011ve,Stecker:2014oxa}.
In this work, we assume that the electron is not also superluminal, which has been independently constrained in, for instance, Ref.~\cite{Stecker:2013jfa}.
The calculation of the decay width $\Gamma = \Gamma(E,\delta)$, where $E$ is the neutrino energy, is presented in Refs.~\cite{Cohen:2011hx,Huo:2011ve} and used, for instance, in Ref.~\cite{Stecker:2014xja} to set a limit on $\delta$.
It is generally possible to set a limit on $\delta$ using any neutrino if we know its energy and propagated distance.
Astrophysical neutrinos, which are neutrinos originating from outside the Solar System, are uniquely useful for this purpose because they arrive at high energies and from long distances, both of which serve to competitively constrain the size of the LIV effect, via the $\delta$ parameter.
Indeed, there have been many previous efforts using astrophysical neutrinos to constrain LIV; see, for example, Refs.~\cite{Laha:2018hsh,Boran:2018ypz,Ellis:2018ogq,Wei:2018ajw,Stecker:2014xja,Stecker:2014oxa,Wang:2020tej,Diaz:2013wia,Diaz:2013wia,IceCube:2017qyp,Arguelles:2015dca,IceCube:2021tdn,Borriello:2013ala,Zhang:2018otj,Huang:2018ham}.

KM3NeT~\cite{KM3Net:2016zxf} is a research infrastructure comprising two detector arrays in the Mediterranean Sea which, among other scientific aims, is being built to detect such astrophysical neutrinos.
Recently ARCA, the larger detector, observed a muon indicative of an ultra-high-energy (UHE) neutrino event, termed KM3-230213A~\cite{KM3NeT:2025npi}, with an estimated neutrino energy
$$E_{\text{UHE}} = 220^{+570}_{-110} \: \text{PeV},$$
which is the highest energy neutrino ever observed to date.
This estimate relies on the assumption that neutrinos of this energy follow a $E^{-2}$ spectrum~\cite{KM3NeT:2025npi}.
In this work, we will use this reported neutrino energy estimate whose physical lower bound is the reconstructed muon energy of $120~\text{PeV}$; this bound still leads to limits with the same order of magnitude.
While the source of KM3-230213A is not yet known, its high energy and likely extragalactic~\cite{KM3NeT:2025npi,KM3NeT:2025aps} ($L\geq 1~\text{Mpc}$) nature already allows us to set a world-leading constraint on $\delta$.

\section{Limit and Discussion}\label{sec:limit_discussion}

\begin{figure}
    \centering
    \includegraphics[width=1.0\linewidth]{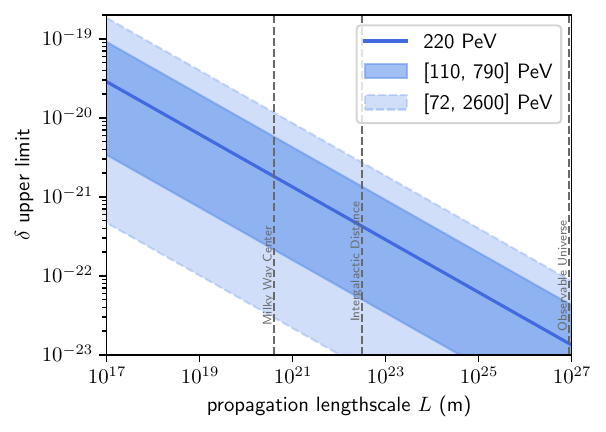}
    \caption{The value of $\delta$ scanning over a wide range of $L$ assuming we hold the energy constant at $E_{\textrm{UHE}}$. The bands correspond to the 68\% and 90\% confidence intervals in the energy estimation of KM3-230213A~\cite{KM3NeT:2025npi}. We also indicate in the vertical dashed lines some lengthscales of interest: the size of the Milky Way, intergalactic distances (1 Mpc), and the size of the observable Universe.}
    \label{fig:limits}
\end{figure}

We set a limit on $\delta$ using the procedure described in Ref.~\cite{Huo:2011ve}. 
First, we calculate $\Gamma$ as given in Ref.~\cite{Cohen:2011hx} and determine a decay length $c_\nu / \Gamma$.
The width has a strong energy dependence, $\Gamma(E,\delta) \propto E^5\delta^3$, which can be reasoned from dimensional analysis as done in Ref.~\cite{Cohen:2011hx} or obtained via a full matrix element calculation as done in Ref.~\cite{Huo:2011ve}.
Secondly, we consider the propagated distance $L$ as ten times the decay length, $L= 10 c_\nu / \Gamma$.
The choice of ten decay lengths is purely conventional, as done in~\cite{Huo:2011ve}, but also conservative - assuming fewer decay lengths traveled for the same $L$ will yield more stringent limits on $\delta$.
Finally, we compute the $\delta$ which is required to produce this $L$ value at a fixed energy $E_{\text{UHE}}$.
The result of this calculation, scanning over a wide range of $L$, is shown in Figure \ref{fig:limits}.
Conservatively taking the minimum distance traveled to be of galactic scale, which means $L\approx4\times 10^{20}\:\textrm{m}$, around the radius of the Milky Way, we can set the limit 
$$\delta < 1.8^{+3.9}_{-1.7} \times 10^{-21},$$
where the range stems from the 68\% confidence interval in the energy measurement~\cite{KM3NeT:2025npi}. 
Given the event direction~\cite{KM3NeT:2025aps}, a more likely scenario would be an intergalactic lengthscale, $L\approx 1~\text{Mpc}$, which results in the limit
$$\delta < 4.2^{+9.2}_{-3.7} \times 10^{-22}.$$

In Table~\ref{tab:other_limits} we show the upper limits on $\delta$, calculated using the same method described above, for other high-energy events and baselines of note. 
In particular, we consider the highest energies and baselines of the IceCube sources NGC~1068~\cite{IceCube:2022der} and TXS~0506+056~\cite{Wang:2020tej,IceCube:2018cha}.
We also show the limit that can be set with IceCube atmospheric neutrinos assuming $(L,E)=(500~\text{km},~100~\text{TeV})$ (e.g., Ref.~\cite{IceCubeCollaboration:2024nle}).

\begin{table}[]
\begin{tabular}{ll}
\hline
Method                                                     & Limit \\ \hline
IceCube atmospheric                                        & $6.2\times10^{-11}$   \\
IceCube NGC 1068                                           & $1.5\times10^{-15}$   \\
IceCube TXS 0506+056                                       & $2.4\times10^{-18}$   \\
Stecker et al. (Ref.~\cite{Stecker:2014xja}) & $5.2\times10^{-21}$   \\
KM3-230213A (conservative)                     & $1.8\times10^{-21}$  \\
KM3-230213A (likely)                     & $4.2\times10^{-22}$ 
\end{tabular}
\caption{A comparison of various limits set with the same method of using 10 decay lengths, with the exception of the limit set by Ref.~\cite{Stecker:2014xja}, which is detailed in that respective work. Limits obtained assuming that the electron is not superluminal.}
\label{tab:other_limits}
\end{table}

Competitive limits of $\mathcal{O}(10^{-18} - 10^{-20})$ have also been obtained with more sophisticated methods such as in Refs.~\cite{Borriello:2013ala,Stecker:2013jfa,Stecker:2014xja}.
For comparison in Table~\ref{tab:other_limits}, we also show the most competitive limit, from Ref.~\cite{Stecker:2014xja}, for which a Monte Carlo approach is used to model spectral distortions in neutrino observations.
This approach is more dependent on the astrophysical flux modelling compared to our method.

As done in  Ref.~\cite{Cohen:2011hx}, there is also the possibility of setting a limit using a defined terminal energy, which is the energy scale after which significant losses do not occur. 
We have confirmed that this method yields a similar limit to within one order of magnitude, $\delta < 2.6\times10^{-22}$ at $E_{\text{UHE}}$ for the likely intergalactic scenario. 

The effect of cosmological redshift can also be considered, which manifests as an effective energy loss that contributes in addition to the pair emission.
If we assume that the neutrino source distribution follows the star formation rate, which peaks at redshift of a few $z$ ~\cite{Behroozi:2012iw}, this represents a $\mathcal{O}(1)$ factor of energy loss and will not have a significant effect that competes with the electron pair emission on intergalactic distances. 
If we assume larger redshifts from even more distant sources, ignoring this additional energy loss effect in the calculation of our $\delta$ limit renders it more conservative.

Finally, the criteria for the primary energy loss mechanism, electron pair emission, is energy-dependent. 
A superluminal neutrino behaves as a particle with an effective mass $E\sqrt{\delta}$; therefore, the energy $E$ of the neutrino must satisfy $E > 2m_e / \sqrt{\delta}$ for pair emission to be possible, where $m_e$ is the electron mass. 
Therefore, as $\delta$ is constrained to be successively smaller, we approach the regime where pair emission may require arbitrarily high neutrino energies, and this mechanism cannot be used to further constrain $\delta$.
For instance, already at $\delta=10^{-22}$ we require $E > 100~\textrm{PeV}$, where $E$ is the energy at which the neutrino decays, necessarily higher than the energy $E_\text{UHE}$ at detection (which is assumed in Figure~\ref{fig:limits}).
Despite this limitation, observable effects, such as distortions in cosmogenic neutrino energy spectra, may still be expected and used to set even tighter limits at ultra-high energies~\cite{Gorham:2012qs}. 

\section{Conclusion}

We report on a new limit on the LIV parameter $\delta$ using KM3-230213A, the most energetic neutrino ever detected to date.
Our result improves upon the current best limits by one order of magnitude, while making minimal and conservative assumptions about the origin of the neutrino.
Given electron pair emission in vacuum as the primary energy loss mechanism, our constraints cannot be significantly improved upon using this method without detecting a neutrino of significantly higher energy, or relieving some of our conservative assumptions.
The competitiveness of our limit highlights the growing role that UHE neutrinos, and neutrino telescopes, can play in testing fundamental symmetries.

\section{Acknowledgements}

The authors acknowledge the financial support of:
KM3NeT-INFRADEV2 project, funded by the European Union Horizon Europe Research and Innovation Programme under grant agreement No 101079679;
Funds for Scientific Research (FRS-FNRS), Francqui foundation, BAEF foundation.
Czech Science Foundation (GAČR 24-12702S);
Agence Nationale de la Recherche (contract ANR-15-CE31-0020), Centre National de la Recherche Scientifique (CNRS), Commission Europ\'eenne (FEDER fund and Marie Curie Program), LabEx UnivEarthS (ANR-10-LABX-0023 and ANR-18-IDEX-0001), Paris \^Ile-de-France Region, Normandy Region (Alpha, Blue-waves and Neptune), France,
The Provence-Alpes-Côte d'Azur Delegation for Research and Innovation (DRARI), the Provence-Alpes-Côte d'Azur region, the Bouches-du-Rhône Departmental Council, the Metropolis of Aix-Marseille Provence and the City of Marseille through the CPER 2021-2027 NEUMED project,
The CNRS Institut National de Physique Nucléaire et de Physique des Particules (IN2P3);
Shota Rustaveli National Science Foundation of Georgia (SRNSFG, FR-22-13708), Georgia;
This work is part of the MuSES project which has received funding from the European Research Council (ERC) under the European Union’s Horizon 2020 Research and Innovation Programme (grant agreement No 101142396);
The General Secretariat of Research and Innovation (GSRI), Greece;
Istituto Nazionale di Fisica Nucleare (INFN) and Ministero dell’Universit{\`a} e della Ricerca (MUR), through PRIN 2022 program (Grant PANTHEON 2022E2J4RK, Next Generation EU) and PON R\&I program (Avviso n. 424 del 28 febbraio 2018, Progetto PACK-PIR01 00021), Italy; IDMAR project Po-Fesr Sicilian Region az. 1.5.1; A. De Benedittis, W. Idrissi Ibnsalih, M. Bendahman, A. Nayerhoda, G. Papalashvili, I. C. Rea, A. Simonelli have been supported by the Italian Ministero dell'Universit{\`a} e della Ricerca (MUR), Progetto CIR01 00021 (Avviso n. 2595 del 24 dicembre 2019); KM3NeT4RR MUR Project National Recovery and Resilience Plan (NRRP), Mission 4 Component 2 Investment 3.1, Funded by the European Union – NextGenerationEU,CUP I57G21000040001, Concession Decree MUR No. n. Prot. 123 del 21/06/2022;
Ministry of Higher Education, Scientific Research and Innovation, Morocco, and the Arab Fund for Economic and Social Development, Kuwait;
Nederlandse organisatie voor Wetenschappelijk Onderzoek (NWO), the Netherlands;
The grant “AstroCeNT: Particle Astrophysics Science and Technology Centre”, carried out within the International Research Agendas programme of the Foundation for Polish Science financed by the European Union under the European Regional Development Fund; The program: “Excellence initiative-research university” for the AGH University in Krakow; The ARTIQ project: UMO-2021/01/2/ST6/00004 and ARTIQ/0004/2021;
Ministry of Research, Innovation and Digitalisation, Romania;
Slovak Research and Development Agency under Contract No. APVV-22-0413; Ministry of Education, Research, Development and Youth of the Slovak Republic;
MCIN for PID2021-124591NB-C41, -C42, -C43 and PDC2023-145913-I00 funded by MCIN/AEI/10.13039/501100011033 and by “ERDF A way of making Europe”, for ASFAE/2022/014 and ASFAE/2022 /023 with funding from the EU NextGenerationEU (PRTR-C17.I01) and Generalitat Valenciana, for Grant AST22\_6.2 with funding from Consejer\'{\i}a de Universidad, Investigaci\'on e Innovaci\'on and Gobierno de Espa\~na and European Union - NextGenerationEU, for CSIC-INFRA23013 and for CNS2023-144099, Generalitat Valenciana for CIDEGENT/2018/034, /2019/043, /2020/049, /2021/23, for CIDEIG/2023/20, for CIPROM/2023/51 and for GRISOLIAP/2021/192 and EU for MSC/101025085, Spain;
Khalifa University internal grants (ESIG-2023-008, RIG-2023-070 and RIG-2024-047), United Arab Emirates;
The European Union's Horizon 2020 Research and Innovation Programme (ChETEC-INFRA - Project no. 101008324);
C. A. Arg\"uelles and N. W. Kamp were supported by the David \& Lucille Packard Foundation; A. Y. Wen was supported by the Natural Sciences and Engineering Research Council of Canada (NSERC), funding reference number PGSD-577971-2023.

\clearpage
\bibliography{main.bib}
\clearpage

\end{document}